\documentclass{emulateapj}
\usepackage{apjfonts}

\submitted{Accepted for Publication in ApJ Letters}

\shorttitle{Intrinsic Brightness Temperatures of AGN Jets}
\shortauthors{Homan et al.}

\begin{document}

\title{\vspace{-0.05in}Intrinsic Brightness Temperatures of AGN Jets}
\author{D. C. Homan\altaffilmark{1}, Y. Y. Kovalev\altaffilmark{2,3}, 
M. L. Lister\altaffilmark{4}, 
E. Ros\altaffilmark{5}, K. I. Kellermann\altaffilmark{6},
M. H. Cohen\altaffilmark{7}, R. C. Vermeulen\altaffilmark{8},
J. A. Zensus\altaffilmark{5}, M. Kadler\altaffilmark{5,9}} 

\altaffiltext{1}{Department of Physics and Astronomy, Denison University, 
Granville, OH  43023; homand@denison.edu}

\altaffiltext{2}{Jansky Fellow, National Radio Astronomy Observatory,
P.O. Box 2, Green Bank, WV 24944; ykovalev@nrao.edu}

\altaffiltext{3}{Astro Space Center of Lebedev Physical Institute,
Profsoyuznaya 84/32, 117997 Moscow, Russia;}

\altaffiltext{4}{Department of Physics, Purdue University, 525 Northwestern
Avenue, West Lafayette, IN 47907; mlister@physics.purdue.edu}

\altaffiltext{5}{Max-Planck-Institut f\"ur Radioastronomie, Auf dem H\"ugel 69,
D-53121 Bonn, Germany; ros@mpifr-bonn.mpg.de,
azensus@mpifr-bonn.mpg.de}

\altaffiltext{6}{National Radio Astronomy Observatory, 520 Edgemont Road,
Charlottesville, VA 22903; kkellerm@nrao.edu}

\altaffiltext{7}{Department of Astronomy, MS 105-24, California
Institute of Technology, Pasadena, CA 91125; mhc@astro.caltech.edu}

\altaffiltext{8}{ASTRON (Netherlands Foundation for Research in Astronomy), 
Postbus 2, NL-7990 AA Dwingeloo, Netherlands; rvermeulen@astron.nl}

\altaffiltext{9}{Present Address: Goddard Space Flight Center, 
Greenbelt, Maryland 20771; mkadler@milkyway.gsfc.nasa.gov}

\begin{abstract}
We present a new method for studying the intrinsic brightness temperatures
of the parsec-scale jet cores of Active Galactic Nuclei (AGN).  Our method uses 
observed superluminal motions and observed brightness temperatures for
a large sample of AGN to constrain the characteristic intrinsic brightness
temperature of the sample as a whole.  
To study changes in intrinsic brightness temperature,
we assume that the Doppler factors of individual jets
are constant in time as justified by their relatively
small changes in observed flux density.
We find that in their median-low  
brightness temperature state, the sources in our sample have a narrow range
of intrinsic brightness temperatures centered on a characteristic
temperature, $T_{int} \simeq 3\times 10^{10}$ K, 
which is close to the value expected for equipartition, when the energy in the radiating 
particles equals the energy stored in the magnetic fields.
However, in their maximum brightness state, we find that sources in our sample 
have a characteristic intrinsic brightness temperature greater than 
$2\times 10^{11}$ K, which is well in excess of the equipartition temperature.  In
this state, we estimate the energy in radiating particles exceeds the energy in the 
magnetic field by a factor of $\sim 10^5$. We suggest that the excess of 
particle energy when sources are in their maximum brightness state is 
due to injection or acceleration of particles at the base of the jet. Our results 
suggest that the common method of estimating jet Doppler factors by using a 
single measurement of observed brightness temperature and/or the assumption of 
equipartition may lead to large scatter or systematic errors in the
derived values. 
\end{abstract}


\section{Introduction}
\label{s:intro}
A key physical property of extragalactic radio jets is the
relationship between the energy stored in the radiating
particles and the energy stored in the magnetic field.  \citet{B59}
originally suggested that these energies were approximately
in balance, or equipartition, as a way of minimizing the 
total energy stored in extended radio lobes, and more recent
work \citep{C05} has indeed shown that extended radio lobes are
at or very near equipartition.  However, it is still unknown whether 
this balance is established at, or close to, the base of the
jet, or if jets begin far out of equipartition and 
only reach equipartition after significant periods of time.

The energy balance at the base of extragalactic radio jets 
can be studied by measuring the brightness temperature of 
jet cores using the Very Long Baseline Array (VLBA).  
The intrinsic brightness temperature, $T_{int}$,
is related to the energy balance via the
following expression \citep{R94}:

\begin{equation}
T_{int} = \eta^{1/8.5}T_{eq}
\end{equation}

\noindent where $\eta = u_p/u_B$ is the ratio of the energy densities
of the radiating particles, $u_p$, to the magnetic field, $u_B$.  $T_{eq} 
\simeq 5 \times 10^{10}$ K is the equipartition brightness temperature defined by 
\citet{R94}, and the exponent of $\eta$ assumes an power-law Lorentz factor
distribution for the radiating particles with index $p = 2.5$ for 
$N_\gamma d\gamma = K\gamma^{-p}d\gamma$. This corresponds to a 
spectral index of $\alpha = -0.75$  ($S\propto\nu^{+\alpha}$) in 
the optically thin parts of the jet. 

The intrinsic brightness temperatures of AGN cores are limited 
either by the inverse-Compton catastrophe to be $T_{int} \lesssim
 10^{12}$ K \citep{KPT69} or
by the possibility that jets are near equipartition to begin with,
so $T_{int} \simeq T_{eq}$ \citep{SG85, S86, R94}.  If AGN cores are 
instead limited by the inverse-Compton catastrophe, they must be very 
far out of equipartition with $\eta \sim 10^{11}$ \citep[e.g.][]{KPT69}.  

Unfortunately, it is very difficult to measure intrinsic
brightness temperatures because AGN jets are highly relativistic and
therefore Doppler boosted.  So with Very Long Baseline Interferometry
(VLBI), we do not measure the $T_{int}$ directly, but rather the observed
brightness temperature\footnote{For this paper we take all observed brightness
temperatures from the point of view of an observer co-moving with the AGN host
galaxy, so we have corrected for redshift.  We also assume that radiation is
emitted isotropically in the co-moving frame, and that the parsec-scale jet 
cores have a flat spectrum ($\alpha = 0$) due to optical depth effects.  Our choice 
of cosmology is $H_0 = 70$ km/s/Mpc, $\Omega_M = 0.3$, and 
$\Omega_\Lambda = 0.7$.} 
$T_{obs} = \delta T_{int}$, where $\delta$ is the Doppler factor of the 
jet, given by

\begin{equation}
\delta = \frac{\sqrt{1-\beta^2}}{1-\beta\cos\theta}
\end{equation}

\noindent where $\beta$ is the speed of the relativistic flow in units of 
the speed of light, and $\theta$ is the angle the jet makes with the line 
of sight.  The speed of the jet is often characterized by its bulk Lorentz
factor, $\Gamma = 1/\sqrt{1-\beta^2}$. 

\begin{center}
\figurenum{1}
\includegraphics[width=2.45in,angle=-90]{f1.eps}
\figcaption[f1.eps]{
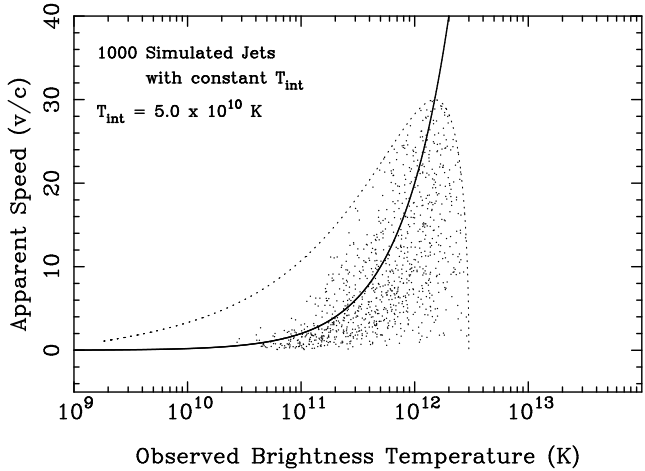
Plot of 1000 fictional sources from a program which simulates
a relativistically beamed population of AGN jets selected 
on the basis of observed flux.  All sources are given the 
same intrinsic brightness temperature.  The solid line 
represents where sources observed at the critical angle
would lie on this plot. The dashed line represents the
possible apparent speeds of a $\Gamma = 30$ source with intrinsic
brightness temperature given by $T_{int}$.
}
\end{center}

Observed brightness temperatures from ground-based VLBI can 
range up to lower limits of $T_{obs} \geq 5 \times 10^{13}$ K 
\citep[e.g.][]{K05}, and observed brightness temperatures in excess 
of $1 \times 10^{14}$ K have been reported with space VLBI \citep[e.g.][]{H04}.  
\citet{H04} find typical observed brightness temperatures in the range of 
$1\times 10^{11}$ K to $1\times 10^{13}$ K from the VSOP 5 GHz survey, 
and \citet{K05} find a similar range from ground-based VLBA measurements 
at 15 GHz. The recent discovery of very nearby scintillation screens
has brought the maximum brightness temperatures deduced from intra-day variability 
(IDV) into the same range as those measured through VLBI even though 
interstellar scintillation is potentially sensitive to much smaller 
angular sizes than the VLBA \citep[and references therein]{L03}.  
      
In this paper, we present a new method for deducing the typical 
intrinsic brightness temperature for parsec-scale AGN cores from 
a population of well studied VLBA jets.  These cores represent the 
optically thick region at the start of the jet as seen in VLBI images. We
combine observed brightness temperature data from \citet{K05} and proper motion 
data from the 2 cm survey 
\citep[e.g.][]{K04}.  In \S{\ref{s:theory}} we 
discuss the theory behind our approach, and in \S{\ref{s:res}} we present
our results and discuss the implications for energy balance between 
particles and magnetic fields at the base of AGN jets.

\section{Methodology}
\label{s:theory}

The basis of our approach is to compare the observed brightness temperatures, $T_{obs}$, 
for a collection of parsec-scale AGN jet cores with the proper motions of their 
jets, $\beta_{app}$. Proper motion, $\beta_{app} = \beta\sin\theta/(1-\beta\cos\theta)$, 
will serve
as a proxy for the jet Doppler factor, $\delta$, which we cannot easily measure.
For the purposes of this analysis, we will assume that the speeds of the fastest jet
components 
are the same as the flow speed through the jet core. We will also assume the 
jets are straight, with no bends between the jet core and the location of jet 
components.  To the degree that these assumptions are incorrect, we should expect 
some scatter in our final results.

To generate a simple model, we make two further assumptions: (1)
that all jets in our sample have the same intrinsic brightness temperature, $T_{int}$,
and (2) that all sources are close to the critical angle, $\theta_c = \arccos\beta$, for the
maximal superluminal motion at a given $\beta$.  Under these strict assumptions, 
$\delta \simeq \beta_{app}$ and $T_{obs} \simeq \beta_{app}T_{int}$, and therefore
$\beta_{app}$ is a simple function of the observed brightness temperature, as 
represented by the solid line in Figure 1.  

\placefigure{1}

Of course, the assumption that all jets are near the critical angle is poor.  
For a flux-limited sample, the majority will lie inside the critical angle where 
the Doppler factor exceeds the Lorentz factor \citep{VC94}. 
The probability distribution for jet orientations in a beamed, 
flux-limited sample has been studied in detail by \citet{VC94} and \citet{LM97}.    
In Figure 1, $\beta_{app}$ is plotted versus $T_{obs}$ from a numerical simulation 
of a flux-limited sample of 1000 AGN jets.  This simulation is based on 
\citet{LM97} and assumes that all 
jets have the same intrinsic brightness temperature $T_{int} = 5 \times 10^{10}$ 
K.  The jet speeds in the population range from $\beta = 
0.05$ to $\Gamma = 30$, and are distributed according to $n(\Gamma) \propto 
\Gamma^{-1.5}$.  These simulated jets are drawn from a 
randomly-oriented parent population 
with an intrinsic
(unbeamed) luminosity function consistent with FR-II radio 
galaxies described in \citet{LM97}.  The luminosity function has a lower cutoff 
at $1 \times 10^{23} \; {\mathrm W \; Hz^{-1}}$, and incorporates pure 
exponential luminosity evolution while maintaining a constant co-moving space 
density out to $z =4$.  

Approximately 75\%
of the sources in Figure 1 fall below and to the right of the solid line
representing the critical angle; these sources are inside the critical angle,
and while they have large Doppler factors, many of them are close enough to
the line of sight to have small proper motions.    The dotted line in Figure 1 
shows the apparent speed for a jet with the maximal Lorentz factor of 
$\Gamma = 30$ in the simulation as a function of $T_{obs}$, corresponding
to a varying jet viewing angle.  The dotted line, therefore, represents an 
upper envelope for the simulation.

The scatter in Figure 1 gives a realistic idea of what we might expect 
for a ``best case'' correlation between $\beta_{app}$ and $T_{obs}$ for real
data under the assumptions that jets are straight, pattern speeds $=$ flow 
speeds, and all jets have the same $T_{int}$.  However, even with the 
scatter induced by selection effects, the trend is clear, particularly the 
lack of sources with {\em both} low observed brightness temperature {\em and} fast 
proper motion.  

\section{Results and Discussion}
\label{s:res}

We have collected multi-epoch brightness temperature measurements for 106 sources
that all have well studied proper motions and at least 5 epochs of brightness 
temperature measurements.  The brightness temperature measurements
are taken from \citet{K05}, and the proper motion measurements are updated from 
\citet{K04} to include additional epochs from the VLBA\footnote{
The Very Long Baseline Array is operated by the National Radio Astronomy Observatory, 
a facility of the National Science Foundation operated under cooperative agreement 
by Associated Universities, Inc.}
2cm Survey\footnote{http://www.cv.nrao.edu/2cmsurvey/} through 2002 (E. Ros et al.,
in prep.).  As described in \citet{K04} we use the fastest proper motion for each
source, and we only consider those motions that are ranked as ``Excellent'' (E) or
``Good'' (G) by the criteria laid out in that paper.  Of the 106 sources plotted
here, 70 are members of the 133 source, flux-limited 
MOJAVE\footnote{http://www.physics.purdue.edu/astro/MOJAVE/} sample \citep{LH05}.  The
remaining 35 sources are members of the 2cm Survey, but do not meet the selection
criteria of the MOJAVE program.  

\placefigure{2}

Figure 2 is a three panel plot of $\beta_{app}$ vs $T_{obs}$.  Figure 2a contains 106
fictional sources from the simulation plotted in Figure 1.  The 106 fictional sources
are selected in a manner analogous to our actual data, with 70 of them chosen at random 
from the 133 brightest sources in the simulation and the remaining 36 chosen at 
random from the rest.  This panel is included to illustrate that selection criteria
similar to our real sample will still produce a plot very similar to Figure 1 with
about 75\% of the points falling below and to the right of the solid line representing
sources at the critical angle.  As in Figure 1, the dotted line in each panel 
is the apparent speed for a jet with a bulk Lorentz factor of 30 
as a function of $\theta$.  If all sources had the same $T_{int}$ and none 
had $\Gamma > 30$, the dotted line would encompass all of the sources.  

Figure 2b is a plot of $\beta_{app}$ versus the maximum observed brightness temperature for
each source in our real sample.  The solid line represents sources at the critical angle which have
$T_{int} = 2 \times 10^{11}$ K, and the same value for $T_{int}$ is used in calculating
the dotted ``envelope.''  This value of $T_{int}$ was chosen so that approximately
75\% of the sources would fall below and to the right of the solid line; thus matching our 
expectation from the simulation.  Figure 2c is a similar plot except that it shows sources
in their median-low state, what we call the ``25\% Median''.  This 25\% median is the median 
of the lowest half of the brightness temperature observations for a given source.  In this 
way, the 25\%-median represents a typical low brightness state for each source.  The 
$\simeq 75$\% line for the 25\%-median values is $T_{int} = 3\times 10^{10}$ K.
A more detailed fit of our data to the distribution of simulation points in Figure 2a is 
beyond the scope of this paper, and we note that other reasonable choices for the simulation 
parameters described in section \ref{s:theory} will give a roughly similar distribution of 
points and are likely to yield fractions between 60\% and 80\% of sources within the 
critical angle.  For a 60\% fraction, we would have deduced $T_{int} = 3\times 10^{11}$ for 
the maximum case and $T_{int} = 4\times 10^{10}$ K for the median-low case.

It is interesting that plotting $\beta_{app}$ versus $T_{obs}$ in figures 2b and 2c 
produces a trend very similar to that seen in the simulation, although with 
more scatter.  There is clearly a lack of sources with {\em both} low observed 
brightness temperature {\em and} fast proper motion.  The fact that the trend is 
still clear here implies that both plots are described by a narrow range 
of intrinsic brightness temperature, centered on the value chosen for the solid 
line in each plot.   If the scatter in the 25\%-median and maximum plots is due to 
intrinsic brightness temperature alone, we estimate the real range of intrinsic 
brightness temperature in each case extends down to about 50\% of the $T_{int}$ for 
the plotted line and up to about twice this value.  Note that this means the 
intrinsic brightness temperature ranges for the 25\%-median and maximum cases do 
not overlap.  Real sources appear to be in a distinctly different 
state when they display their maximum brightness temperatures.  

We note that the apparent increase in brightness temperature between the 25\%-median 
and maximum states cannot be attributed to changes in the jet Doppler factors, which 
we have assumed to be constant in time, because the necessary large changes in Doppler 
factor would produce much larger changes 
in the flux densities of the jet cores 
($\propto \delta^2$) which are not observed.

\begin{center}
\figurenum{2}
\begin{center}
\includegraphics[width=3.1in,angle=0]{f2.eps}
\end{center}
\figcaption[f2.eps]{
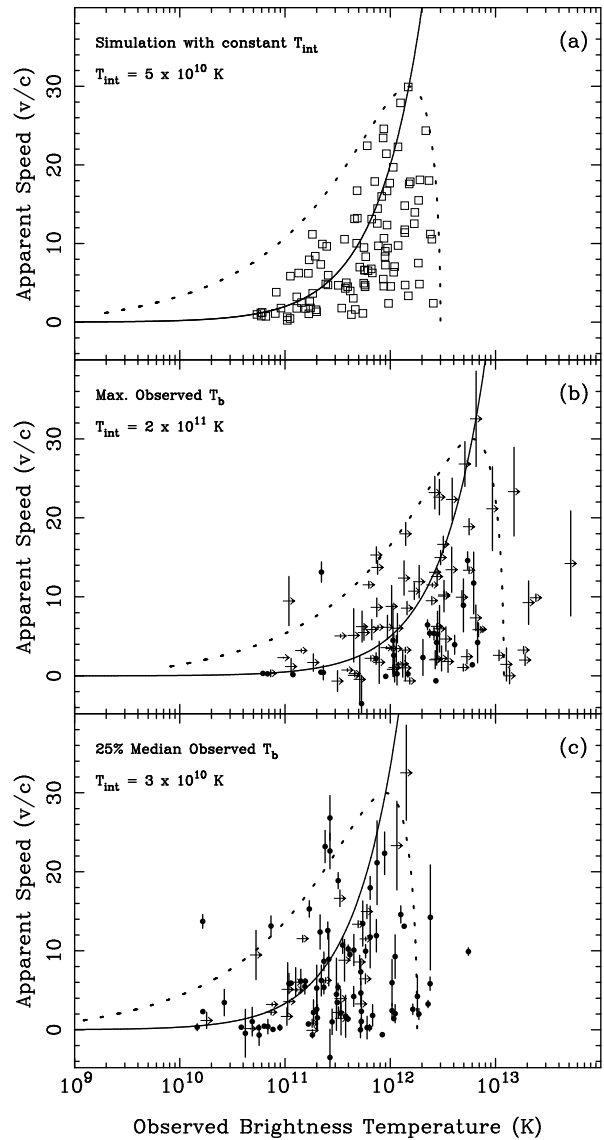
Plots of apparent motion versus observed brightness temperature.  As
described in the text, panel (a) contains 106 sources from the
simulation illustrated in Figure 1, choosen to simulate the selection
characteristics of our real data.  Panels (b) and (c) plot our
real data.  Lower limits are indicated by arrows and solid dots represent
measurements.  The solid line represents sources observed at the 
critical angle that have the intrinsic brightness temperature indicated 
in the upper left hand corner of the panel. The dashed line represents the
possible apparent speeds of a $\Gamma = 30$ source with intrinsic
brightness temperature given by $T_{int}$.
}
\end{center}

The maximum plot contains more lower limits than measurements, so the characteristic
value of $T_{int}$ should really be taken as a lower limit on the characteristic value.
Therefore, $T_{int} > 2 \times 10^{11}$ K when sources display their maximum brightness
temperature.  By equation 1, this requires sources in that state to be well out of 
equipartition, with $\eta \sim  10^5$ or greater.  On the other hand, the characteristic
value for the median-low state is much closer to the equipartition brightness temperature
defined by \citet{R94}, and there are far fewer limits in that plot\footnote{We note
that we tried to examine the pure median case, as representative of the average state
of a source; however, even the pure median case had a large number of limits (about half), 
and we were therefore not able to do a robust comparison with the maximum case.  The 75\% line
for the pure median case was at $6\times 10^{10}K$.}, so it seems 
reasonable to take the characteristic median-low intrinsic brightness temperature 
to be $T_{int} \simeq 3\times10^{10}$ K.   

Our data indicate that AGN jet cores are indeed near equipartition in their median-low state;
however they go well out of equipartition in their maximum brightness temperature state.
In their maximum state they have $\gtrsim 10^5$ times the energy in their radiating 
particles 
as they have in the magnetic fields.  Such a large imbalance may be the result of 
injection or acceleration of particles at the base of the jet, resulting in a transient
state that is in excess of the equilibrium equipartition brightness temperature.  
For $\eta \sim 10^5$, equation 5 from \citet{R94} predicts a particle number density 
which has increased by a factor of 400 and a magnetic field strength which has 
decreased by a factor of 15 compared to the equilibrium value.  We emphase that these
values are only approximate as the precise balance between particle and field energy
is a sensitive function of brightness temperature (see equation 1).  What is robust, is 
that the jets in our sample have a narrow range of intrinsic brightness temperatures in their
median-low state, and the brightness temperatures in their maximum state are nearly an
order of magnitude (or more) larger, indicating a much higher particle to field
energy ratio.

It is interesting to note that the narrow range of intrinsic brightness temperatures 
when sources are in their median-low state means we can derive Doppler factors for most 
sources which are good to a factor of two or better by assuming $T_{int} = 3\times10^{10}$ K
for the 25\%-median observed value.  Then 
$\delta = T_{obs(25\%-median)}/3\times10^{10}$ K.
This is very similar to the equipartition Doppler factor suggested by \citet{SG85} and 
\citet{R94}; however, by using the 25\%-median observed brightness temperature we 
get a much more 
reliable estimate for the Doppler factor.  Doppler factors derived from single
epoch brightness temperature measurments \citep[e.g.][]{Z02} are likely to have a large 
degree of scatter as it is clear from this work that the intrinsic brightness temperatures
of sources can vary by an order of magnitude or more. Additionally, Doppler factors
derived from outburst events \citep[e.g.][]{LV99} are likely to be systematically overestimated
if it is assumed that sources are near equipartition during these events.

\acknowledgments

We thank the anonymous referee for helping us to clarify several important points, and A. Lobanov
for helpful discussions.  This research was supported by an award from Research Corportation and 
NSF grant 0406923-AST.  A. Z. acknowledges support from the Max-Planck-Award for International 
Research Collaboration.  M. K. was supported through a stipend from the International Max 
Planck Research School for Radio and Infrared Astronomy at the University of Bonn.



\end{document}